\documentclass[twocolumn,epsfig]{mn2e}
\usepackage{color}
\usepackage{graphicx,graphics}
\usepackage{epstopdf}
\newcommand{\mincir}{\raise
-2.truept\hbox{\rlap{\hbox{$\sim$}}\raise5.truept
\hbox{$<$}\ }}
\newcommand{\magcir}{\raise
-2.truept\hbox{\rlap{\hbox{$\sim$}}\raise5.truept
\hbox{$>$}\ }}
\newcommand{\minmag}{\raise-2.truept\hbox{\rlap{\hbox{$<$}}\raise
6.truept\hbox{$>$}\ }}
\newcommand{\be}{\begin{equation}}
\newcommand{\ee}{\end{equation}}
\newcommand{\ba}{\begin{eqnarray}}
\newcommand{\ea}{\end{eqnarray}}
\newcommand{\brr}{\begin{array}}

\newcommand{\err}{\end{array}}
\newcommand{\bc}{\begin{center}}
\newcommand{\ec}{\end{center}}

\title[Tidal flares]{PS1-10jh - a tidal disruption event with an extremely low disk temperature}
\author[Matias Montesinos Armijo and J.A. de Freitas Pacheco]{Matias Montesinos Armijo$^1$ and J.A.~de Freitas Pacheco$^{2,3}$\\
$^1$Departamento de Astronom\'ia y Astrof\'isica, Pontificia Universidad Cat\'olica de Chile, Santiago, Chile\\
$^2$Departamento de F\'isica, Universidade Federal do Espirito Santo, Vitoria, Brazil\\
$^3$Universit\'e de Nice-Sophia Antipolis, Observatoire de la C\^ote d'Azur, BP 4229, F-06304, Nice Cedex 4, France\\
emails: mmontesi@astro.puc.cl; pacheco@oca.eu}

\begin{document}

\date{\today}

\maketitle


\begin{abstract}

The cooler than expected optical-UV transient PS1-10jh detected by the Pan-STARRS1 survey is probably related to 
a tidal disruption event in which a He-rich stellar core remnant is implied. The evolution of bound debris during the
disk phase is studied by solving the hydrodynamic equations. The model provides a good fit either of the raising part of the 
light curve in the bands $g_{P1}$, $r_{P1}$, and $i_{P1}$ or in the early decay. The parameters characterizing this
optimized model are the mass of the central black hole, i.e., $6.3\times 10^6$ M$_\odot$ and the critical Reynolds 
number ${\cal R} = 10^4$ that fixes the viscosity and the accretion timescale. Such a high value of ${\cal R}$
explains the low disk temperature and the consequent absence of X-ray emission. The 
predicted bolometric peak luminosity is about $10^{45}$ $\rm erg s^{-1}$ and the predicted total radiated
energy is about $E_{rad}=2.67\times 10^{51}$ erg.
 
\end{abstract}


\section{Introduction}

Different flare events have been detected either in X-rays (Bade et al. 1996; Komossa \& Bade 1999; Donley et al. 2002; Esquej 
et al. 2007, 2008; Cappelluti et al. 2009; Maksym et al. 2010) or in the UV region (Gezari et al. 2006, 2009). These sudden 
variations in the electromagnetic emission of ``quiet" galaxies have been interpreted as being the consequence 
of the tidal disruption of a star that has passed close to a supermassive black hole (SMBH) living in a ``dormant" state in the 
center of its host galaxy. For a near parabolic orbit in which the periapse distance $R_p$ is shorter 
than the tidal radius $R_t$, tidal disruption is likely to occur and 
two distinct phases in the dynamical evolution of the debris can be distinguished. In the first, the ``fallback" phase,
nearly half of the debris is in gas streams having approximately ballistic trajectories, which converge and collide at periapse, 
dissipating energy through shocks. Typically, for a black hole of mass $M_{bh} \sim 10^7$~M$_\odot$ and a disrupted star of
$m_* \sim 1 M_\odot$ the released energy in this phase is about $10^{51}$ erg. Since the dissipated energy in these shocks 
scales as $E_{sh}\propto M_{bh}^{2/3}m_*^{2/3}$, in our case, as we shall see below, this is expected to be one order of magnitude
less than the ``disk-phase.  As the bound material converges at periapse and dissipates energy,
a small accretion disk is formed and the second phase initiates as soon as the black hole begins to accrete mass, which
coincides with the instant of maximum in the light curve. This is dictated by the viscous mechanism that 
controls the angular momentum transfer, fixing the
timescale in which the material is conveyed from the outer regions of the disk to the last stable orbit. In 
this ``disk"-phase, most of the flare energy is radiated. 

Non-steady accretion disk models have recently been built by Montesinos \& de Freitas Pacheco (2011a, b, hereafter MP11A 
and MP11B) by solving 
numerically the hydrodynamic and the energy transfer equations. These models are able to explain the 
sudden variation of the X-ray emission observed in non-active galaxies that are supposed to be related to tidal 
events like those observed in NGC 3599 or IC 3599. More recently, Gezari et al. (2012) reported an UV-optical flare
that has occurred in a non-active galaxy at $z = 0.1696$ and which was probably originated in a tidal disruption event. This
flare event discovered in the Pan-STARRS1 survey
displays some particular characteristics like: a) the lack of hydrogen lines in the spectra and the presence of a broad
HeII$\lambda$4686 \AA~ emission line suggest that the disrupted star lost its outer envelope and can 
probably be identified with the He-rich core of an ancient red giant; b) no detectable X-rays emission by 
Chandra; c) from the analysis of the continuum emission, the estimated disk temperature 
is considerably lower than those attributed to disks associated usually with X-ray flare events.

We report here the results obtained from a series of models computed with our code, aiming to reproduce the observed
light curve in different wavelengths and to derive the main
parameters characterizing the event.

\section{The model}  

The hydrodynamic equations describing the disk evolution can be
found in MP11A and MP11B respectively but, in a few words, we 
recall here the basic features. The code is based on an Eulerian formalism using a
finite difference method of second-order, according to the Van Leer upwind algorithm on a staggered mesh. Since the disks 
considered here are quite small, we have adopted an integration
grid of 256 ring sectors instead of the original 1024 rings adopted in MP11A. The inner radius of the grid coincides with
the last stable orbit while the external radius is defined by the
tidal radius $R_t = \mu R_*(M_{bh}/m_*)^{1/3}$, where $m_*$ and
$R_*$ are respectively the mass and radius of the disrupted star
and $M_{bh}$ is the black hole mass. Notice that generally in the
literature the parameter $\mu$ is taken to be equal to unity. In fact, the work by Luminet \& Carter (1986) suggests that
$\mu$ = 2.4, which corresponds to the Roche limit for a fluid body in a circular orbit. More recent investigations of the
fly-by problem for a viscous fluid body in a parabolic orbit suggest an effective Roche limit ($R_p = R_t$) with $\mu = 1.69$
(Sridhar \& Tremaine 1992; Kosovichev \& Novikov 1992). In particular, the latter authors considered the case in which the
disturbing body is a massive black hole as in the present case. Thus, the value $\mu = 1.69$ will be adopted in our grid
of models. This implies that in our models the most 
bound material returns to periapse after a timescale $t_{min}$ that is about 4.8 times higher than the case $\mu$=1. As 
we shall see later, this explains why Gezari et al. (2012) concluded that the
disruption occurred only $\sim$ 76 days before maximum light while our preferred model indicates a higher timescale, namely,
1.24 years. The original MP11A boundary conditions were also
modified to allow an inflow of matter at the external ring, corresponding to the material that circularizes after shocks
between converging streams.

Also as in MP11A, it was assumed that the gravitational field of the
black hole is given by the approximate potential of Paczynski-Wiita (Paczynski \& Wiita 1980) that gives correctly the position
of the last stable orbit for the Schwarzschild geometry.

It should be emphasized that in the present model the angular
momentum transfer is not described by the so-called 
``$\alpha$"-model introduced by Shakura \& Sunyaev (1973). The viscosity coefficient is that given by the approach of de Freitas
Pacheco \& Steiner (1976), i.e., $\eta = 2\pi rV_{\phi}/{\cal R}$, where $r$ is the radial 
distance to center of the disk, $V_{\phi}$ is the azimuthal velocity of the debris at that 
distance and ${\cal R}$ is the critical Reynolds number characterizing the flow.

Another important aspect concerns the fallback rate. According 
to early investigations (Rees 1988; Evans \& Kochanek 1989), in 
a first approximation, the fallback rate $\dot R_{fb}$ is given by the relation
\begin{equation}
\dot R_{fb} \simeq \frac{1}{3}\frac{m_*}{t_{min}}
\left(\frac{t_{min}}{t}\right)^{5/3},
\label{fb1}
\end{equation}
where
\begin{equation}
t_{min}=\frac{\pi}{\sqrt{2}}\frac{R^3_p}{(GM_{bh}R^3_*)^{1/2}}.
\label{tmin}
\end{equation}
However, more detailed studies (Lodato et al. 2009) based on 
simulations indicate deviations from eq.\ref{fb1} at early 
phases of the fallback process. In fact, only in the late
evolutionary stages the fallback rate varies as $t^{-5/3}$. As 
in MP11B, we have attempted to include these deviations by 
modifying eq.\ref{fb1} as
\begin{equation}
\dot R_{fb} = A\frac{(t/t_{min})^{1/3}}{\left[a+(t/t_{min})^2\right]}.
\label{fb2}
\end{equation}
In the above equation $A$ is a normalization constant (see
MP11A for details) and 
$t_{min}$ is still given by eq.\ref{tmin}. Notice that the 
dimensionless parameter $a$ permits to control the instant at
which the maximum fallback rate occurs.

\section{Results}

A series of models were computed in which the different parameters
were varied, searching for a best representation
of data, namely, the light curves of the flare in different 
wavelengths. In all models, we assumed that the disrupted star
is a He-rich core of 0.23 $\rm M_\odot$ and a radius of 0.33 
$R_\odot$ as in Gezari et al. (2012). Our preferred model is characterized by a black hole of mass 
$6.3\times 10^6$ $\rm M_\odot$,
a fallback parameter $a = 0.06$ and a critical Reynolds number
${\cal R} = 10^4$. Notice that this value is higher than those
characterizing our previous models (MP11B) but is necessary to
get a ``colder" disk. A higher value of ${\cal R}$ decreases
the local dissipation of kinetic energy, producing lower 
temperatures and consequently, increasing the energy inflow by advection.

\begin{figure}
\begin{center}
\includegraphics[height=6cm,width=8cm]{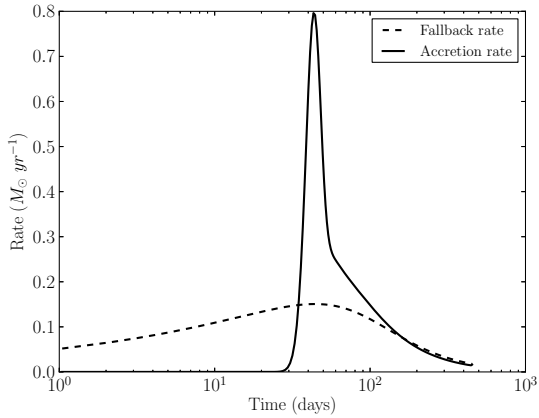}
\end{center}
\vfill
\vspace{0.2cm}
\caption{Evolution of the fallback rate (dashed curve) compared with the evolution of
accretion rate by the black hole (solid curve).}
\label{rate}
\end{figure}

Contrary what is generally assumed in the literature, the 
beginning of the accretion process by the black hole does
not coincide with the beginning of the circularization process 
or, in other words, with the beginning of the formation of the
accretion disk. This is simply because it takes a finite time
for the debris to be transported from the outer regions down to
the inner region (last stable orbit) due to the angular momentum
transfer mechanism. This is clearly seen in fig.1 where the evolution of the fallback rate is compared with the evolution of
the accretion rate by the black hole. The very initial peak
in the accretion rate occurs (for our preferred model) about
44.8 days after the onset of the circularization process and coincides also with the maximum 
in the light curve. The mass of the gravitationally bound debris is about 0.115 $\rm M_\odot$, corresponding to
a half of the mass of the disrupted star, but a small
fraction (about 15\%) is not accreted by the black hole, escaping from the system and taking away 
the angular momentum stored in the disk.

The continuous emission of the flare is generally fitted by a black-body
distribution characterized by an effective temperature. In reality, the local effective 
temperature varies along the disk surface with a radial profile that evolves in time. MP11B defined a 
suitable mean effective temperature, which corresponds
to an effective energy flux that multiplied by the total disk surface gives the luminosity. Such a 
mean effective temperature
is computed from the relation
\begin{equation}
<T^4> = \frac{\int^{R_p}_{R_{lso}}rT^4(r)dr}{\int^{R_p}_{R_{lso}}rdr}.
\end{equation}

\begin{figure}
\begin{center}
\includegraphics[height=6cm,width=8cm]{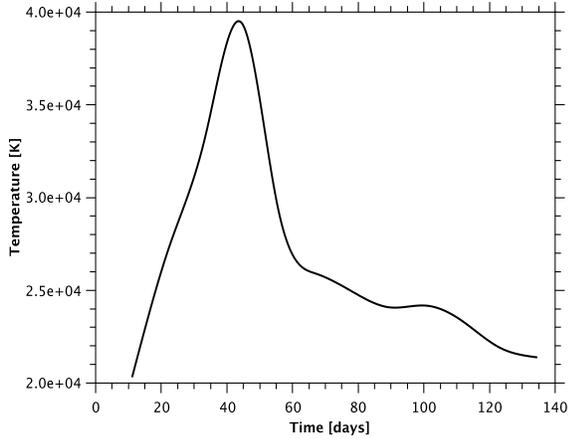}
\end{center}
\vfill
\vspace{0.2cm}
\caption{Evolution of the average effective temperature of the disk.}
\label{temp}
\end{figure}

The evolution of the mean effective temperature is shown in fig.2. The maximum value ($\sim$ 39400 K) coincides
with maximum light (and maximum accretion rate by the black hole).
After approximately 25 days the mean effective temperature of the disk drops to about 25000 K
and then decreases more or less steadly at a rate of $\sim 60 K/day$.
The model predicts at maximum light and near the central region a ``true" temperature around 65000 K. One month after maximum,
in the inner region, the temperature has decreased to about 31000 K while in the outskirts of the disk it remains at a 
value around 20000 K. Figure 3 shows different snapshots of the temperature profile before, at and after maximum light.

\begin{figure}
\begin{center}
\includegraphics[height=6cm,width=8cm]{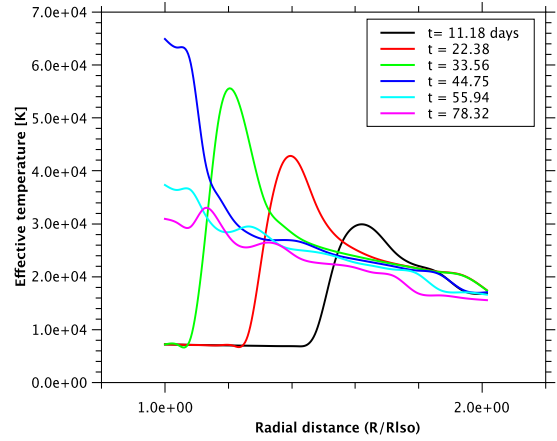}
\end{center}
\vfill
\vspace{0.2cm}
\caption{Snapshots of the temperature profile of the disk. Labels indicate the corresponding instants of time
after the beginning of the circularization process.}
\label{tempprofile}
\end{figure}

In figure 4, the bolometric light curve is shown. As already mentioned, before maximum light the black hole
is not yet accreting matter and the radiation comes from viscous dissipation in the disk that is still being
formed. Integrating the bolometric luminosity over time allows an estimation for the
total radiated energy, i.e., $E_{rad}=2.67\times 10^{51}$ erg, in agreement with the lower limit estimated by Gezari et al. (2012), 
based on a black-body spectrum characterized by a temperature $T_{BB} \geq$ 30 000  K.
Taking into account the total amount of mass accreted by the black hole, the mean
efficiency for the energy conversion is $\eta = E_{rad}/(M c^2)\simeq 0.015$.

\begin{figure}
\begin{center}
\includegraphics[height=6cm,width=8cm]{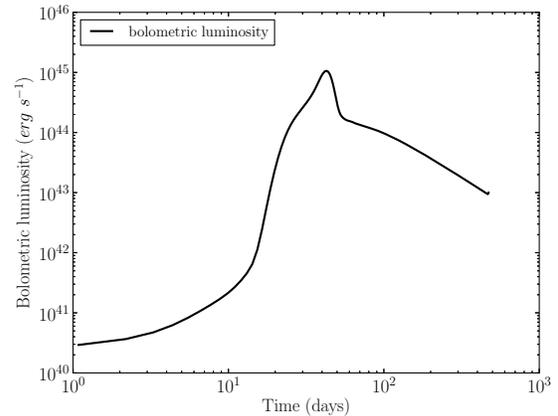}
\end{center}
\vfill
\vspace{0.2cm}
\caption{Evolution of the bolometric luminosity.}
\label{temp}
\end{figure}

\begin{figure}
\begin{center}
\includegraphics[height=6cm,width=8cm]{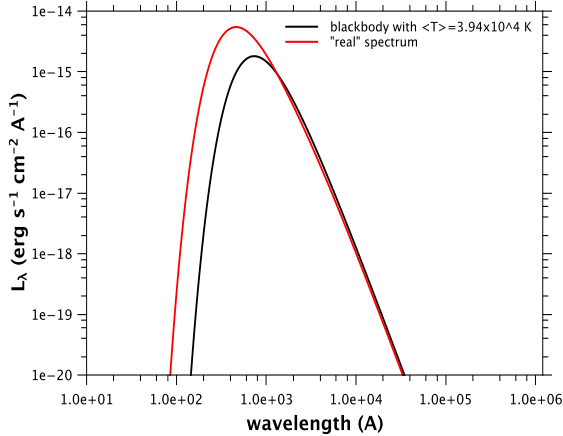}
\end{center}
\vfill
\vspace{0.2cm}
\caption{Comparison between the ``real'' spectrum at the peak instant with the 
equivalent blackbody spectra characterized by the mean temperature of the disk at the peak.}
\label{spectrum}
\end{figure}

The resulting spectrum at maximum is shown in fig. 5 in comparison with the expected 
spectrum of a black body characterized by the mean effective temperature. As MP11B have shown, the black 
body approximation underestimates the emitted
flux at short wavelengths ($\lambda \leq$ 1000 \AA) but predicts correctly fluxes at longer wavelengths. 

\begin{figure}
\begin{center}
\includegraphics[height=13cm,width=7.5cm]{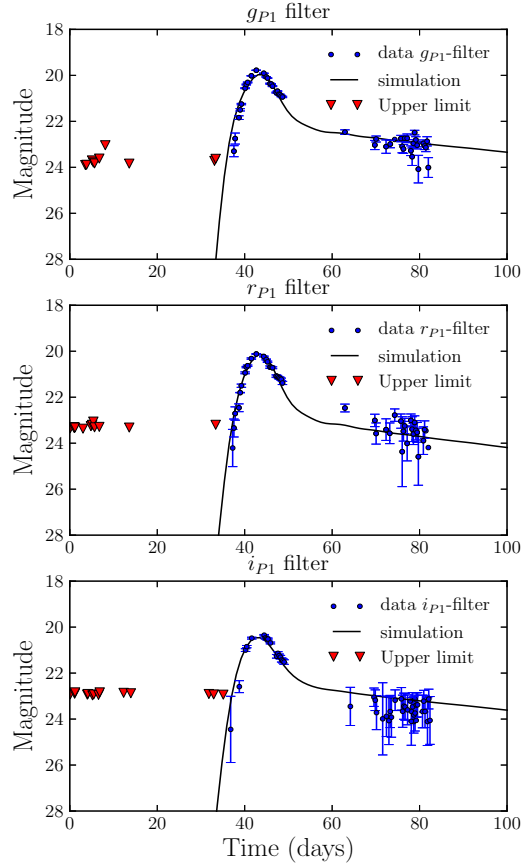}
\end{center}
\vfill
\vspace{0.2cm}
\caption{Simulated light curves (solid continuous curves) compared with data from Gezari et al. (2012). Upper 
panel correspond to $g_{P1}$ filter, middle panel to $r_{P1}$, and the lower one to 
$i_{P1}$.}
\label{spectrum}
\end{figure}

The light curve in filters $g_{P1}$, $r_{P1}$ and $i_{P1}$ are shown respectively in the upper, middle and lower panels of
figure 6. Data points are those given by Gezari et al. (2012).
According to these authors, who have adopted the models by
Lodato et al. (2009), systematic differences are present between
data and theoretical light curves during the early rise and the
late decay. The present model explains quite well both the raising part of the light curve and the 
early decay phase. The former is produced while the disk is still being formed and the black hole is not 
yet accreting mass. The early decay, just after maximum does not vary as $t^{-5/3}$ as usually claimed but 
varies faster, decreasing by about one magnitude per week. The late evolution of the theoretical light curve 
displays the aforementioned behaviour, i.e., $L\propto t^{-5/3}$ although data in the $i_{P1}$ filter (lower panel) 
$\sim$ 40 days after maximum are slightly below the predicted values but still consistent taking into account the
observational errors.
 
\section{Summary}

The transient event discovered on 2010 May 28 by the Pan-STARRS1
(PS1) Medium Deep Survey was probably originated in a tidal disruption event. According to 
Gezari et al. (2012) the disrupted star was a helium-rich core resulting from the evolution of a red giant. We have 
used our hydrodynamic code (MP11A, MP11B) to
compute the radiation of the bound debris during the phase in which a small accretion disk is formed and during which most of
the flare energy is produced.

Our preferred model requires that the central black hole has a
mass of $6.3\times 10^6$ M$_{\odot}$ almost a factor 2.3 higher
than that estimated by Gezari et al. (2012) but in good agreement
with the relation between the black hole mass and the stellar mass of the host galaxy.

No significant X-ray emission related to the flare was detected,
consistent with the temperature of $\sim$ 35 000 K estimated
by Gezari et al. (2012) that is necessary to produce enough UV photons to ionize helium. Our model predicts at maximum and near
the inner region of the disk a temperature of 65 000 K, which is a consequence
of a high critical Reynolds number (${\cal R}=10^4$) characterizing the turbulent state of the gas. This is about 20 times
the ${\cal R}$ values defining our previous models (MP11B), in which an important X-ray emission is
present. Notice that in our model the heating rate per unit volume $\varepsilon$ due to viscosity is given by
the equation
\begin{equation}
\varepsilon =\frac{2\pi}{{\cal R}}\rho r^2\Omega^3\left(\frac{dlg\Omega}{dlg r}\right)^2
\end{equation} 
that shows why a higher critical Reynolds number or a lower viscosity decreases the heating rate and, consequently,
the disk temperature.

The most important difference with the analysis by Gezari et al.
(2012) concerns the chronology of the event, which results from the adopted value for the dimensionless parameter $\mu$ defining
the tidal radius. If $\mu\simeq$ 1.69 and not one, then the disruption of the star occurred about 1.24 yr before maximum light
and the beginning of the circularization process, about 44.7 days before the peak. We emphasize once more that the beginning of 
circularization doesn't coincide with the beginning of the accretion process by the black hole.

Contrary to the models adopted by Gezari et al. (2012), the present model provides a good representation of the initial
raise of the light curve, when the black hole is not yet
accreting and the disk is being filling up, as well as of
the initial light decay, which evolves faster than 
$d\log L/d\log t = 0.6$.

Detailed analyses of tidal disruption events provide an excelent
tool to study dormant black holes and the physics of non-steady
accretion disk, in particular of the mechanisms reponsible
for the angular momentum transfer and energy dissipation.

\section*{Acknowledgments}

MMA acknowledges the support from FONDECYT through grant No. 3120101, and Basal (PFB0609). The authors are grateful to the referee
for her/his valuable comments which have improved considerably this paper.

\end{document}